\documentclass[preprint,floats,prl,aps]{revtex4}

\usepackage[dvips]{graphicx}
\usepackage{amsfonts}
\usepackage{amsmath}
\usepackage{amssymb}
\usepackage{exscale}

\begin{document}

\title{
		Unveiling a two-dimensional electron gas with universal subbands at the surface of SrTiO$_3$
	  }

\author{A.~F.~Santander-Syro$^{1,2,\star}$, 
		O.~Copie$^{3,4}$,
		T.~Kondo$^{5}$, 
		F.~Fortuna$^{1}$,
		S.~Pailh\`es$^{6}$,
		R.~Weht$^{7,8}$,
		X.~G.~Qiu$^{9}$,
		F.~Bertran$^{10}$,
		A.~Nicolaou$^{10}$,
		A. Taleb-Ibrahimi$^{10}$,
		P. Le F\`evre$^{10}$,
		G.~Herranz$^{11}$,
		M.~Bibes$^{3}$,
		Y.~Apertet$^{12}$,
		P.~Lecoeur$^{12}$,
		M.~J.~Rozenberg$^{13,14}$
		\& A.~Barth\'el\'emy$^{3}$. 
		}

\address{\footnotesize{
		 $^1$CSNSM, Universit\'e Paris-Sud and CNRS/IN2P3
		 B\^atiments 104 et 108, 91405 Orsay cedex, France.
 	   	 \\
		 $^2$Laboratoire Photons Et Mati\`ere, UPR-5 CNRS, ESPCI,
 	   	 10 rue Vauquelin, 75231 Paris cedex 5, France.
		 \\
		 $^3$Unit\'e Mixte de Physique CNRS/Thales, Campus de l'Ecole Polytechnique, 1 Av. A. Fresnel, 
		 91767 Palaiseau, France and Universit\'e Paris-Sud, 91405 Orsay, France.
		 \\
		 $^4$Universit\"at W\"urzburg, Experimentelle Physik VII, 97074 W\"urzburg, Germany.
		 \\
		 $^5$Ames Laboratory and Department of Physics and Astronomy, Iowa State University,
		 Ames, IA 50011.
		 \\
		 $^6$Laboratoire L\'eon Brillouin, CEA-CNRS, CEA-Saclay, 91191 Gif-sur-Yvette, France.
		 \\
		 $^7$Gerencia de Investigaci\'on y Aplicaciones, CNEA, 1650 San Mart\'in, Argentina.
		 \\
		 $^8$Instituto S\'abato, Universidad Nacional de San Mart\'in, CNEA, 1650 San Mart\'in, Argentina. 
		 \\
		 $^9$Institute  of Physics and National Laboratory for Condensed Matter Physics,
		 Chinese Academy of Sciences, Zhongguancun nansanjie 8, Beijing 100190, China.
		 \\
		 $^{10}$Synchrotron SOLEIL, L'Orme des Merisiers, Saint-Aubin-BP48, 91192 Gif-sur-Yvette, France.
		 \\
		 $^{11}$Institut de Ci\`encia de Materials de Barcelona, CSIC, Campus de la UAB, 
		 08193 Bellaterra, Catalunya, Spain.
		 \\
		 $^{12}$Institut d'Electronique Fondamentale, Universit\'e Paris-Sud, B\^atiment 220, 91405 Orsay, France.
		 \\
		 $^{13}$Laboratoire de Physique des Solides, Universit\'e Paris-Sud, B\^atiment 510, 91405 Orsay, France.
		 \\
		 $^{14}$Departamento de F\'isica, FCEN-UBA, Ciudad Universitaria, Pab.~1, Buenos Aires (1428), Argentina.
		 \\
		 $^{\star}$e-mail:~andres.santander@csnsm.in2p3.fr  
		}
		}
\maketitle

{\bf
	Similar to silicon that is the basis of conventional electronics, strontium titanate (SrTiO$\mathbf{_3}$) 
	is the bedrock of the emerging field of oxide electronics~\cite{Ramirez-NV-QHE-ZnO-2007, Manhart-OxideNanoElectronics-2009}.
	SrTiO$_3$ is the preferred template to create exotic two-dimensional (2D) phases
	of electron matter at oxide interfaces~\cite{Ohtomo-Wang-2004, Ahn-2003, Ohtomo-2002},
	exhibiting metal-insulator transitions~\cite{Thiel-TunableQ2DEG-Science-2006, Cen-NanoPatterning-LAO-STO-2008}, 
    superconductivity~\cite{Reyren-SC-interface-LAO-STO-Science-2007, Ueno-Electric-Field-Induced-SC-Plastic-STO-2008}, 
    or large negative magnetoresistance~\cite{Brinkman-MagneticInterface-LAO-STO-NatMat-2007}.
    However, the physical nature of the electronic structure underlying these 2D electron gases (2DEGs) remains elusive, 
    although its determination is crucial to understand their remarkable 
    properties~\cite{Okamoto-Millis-2004, Popovic-LAO-STO-FS-ElectronicStructure-PRL2008}.
	Here we show, using angle-resolved photoemission spectroscopy (ARPES), that there is a 
	highly metallic universal 2DEG at the vacuum-cleaved surface of SrTiO$_3$, 
	independent of bulk carrier densities over more than seven decades, including 
	the undoped insulating material.
	This 2DEG is confined within a region of $\mathbf{\sim 5}$~unit cells 
	with a sheet carrier density of $\mathbf{\sim 0.35}$ electrons per $a^2$ ($a$ is the cubic lattice parameter).
	We unveil a remarkable electronic structure consisting on multiple subbands of heavy and light electrons.
	The similarity of this 2DEG with those reported in SrTiO$_3$-based
	heterostructures~\cite{Thiel-TunableQ2DEG-Science-2006, Reyren-SC-interface-LAO-STO-Science-2007, 
	Thales-CAFM-Size2DEG-2008} 
	and field-effect transistors~\cite{Ueno-Electric-Field-Induced-SC-Plastic-STO-2008, Tokura-Parylene-FET-2009}
	suggests that different forms of electron confinement at the surface of SrTiO$_3$ lead to essentially 
	the same 2DEG. Our discovery provides a model system for the study of the electronic structure 
	of 2DEGs in SrTiO$_3$-based devices, and a novel route to generate 2DEGs at surfaces of transition-metal oxides.  
}
\\
\\
In the cubic phase, stoichiometric SrTiO$_3$ has an empty $t_{2g}$ conduction manifold composed of 
three dispersive three-dimensional (3D) bands that are degenerate at the $\Gamma$ point~\cite{Mattheiss-1972-STO}. 
As schematically shown in figure~1(a), its band structure along one direction, say $k_y$, 
consists of a weakly dispersive (heavy mass) band and a pair of degenerate strongly dispersive (light mass) bands. 
They arise, respectively, from the small and large overlaps along $y$ 
of neighbouring Ti $3d_{xz}$, $3d_{xy}$ and $3d_{yz}$ orbitals,
as depicted in figure~1(b) for the $3d_{xy}$ orbitals.
Thus, the $d_{xy}$-like band will be light in the $xy$ plane and heavy along $z$, 
while the $d_{yz}$ band will be heavy along $x$ and light in the $yz$ plane, 
and the $d_{xz}$ band will be heavy along $y$ and light in the $xz$ plane. The resulting three Fermi surface sheets, 
projected on the $xy$ plane of the sample surface, are shown in figure~1(c) for several bulk dopings.
%

ARPES is a powerful technique to probe the electronic structure of materials~\cite{Hufner-Book}. 
Previous ARPES experiments addressed the bulk electronic structure of doped SrTiO$_3$, 
revealing dispersing quasiparticle peaks, in-gap features, 
and discussing possible polaronic effects near the Fermi energy ($E_F$)~\cite{Fujimori-1992-LaSrTiO3, Aiura-2002,
Shin-SoftX-ARPES-STO-IGS, Takizawa-O2pBands-2009, Meevasana-ARPES-STO-2010, Rotenberg-ARPES-STO-CubicTetra-2010}.
As we shall see, our ARPES data do not conform to the expected lightly-doped bulk band picture, 
but reveal that a novel 2DEG with a complex subband structure, schematized in Figs.~1(d,e),
is realized at the low-temperature vacuum-cleaved surface of this material.

Crucial to our discoveries is the comparison of samples with very different bulk carrier densities 
of $n_{3D} \lesssim 10^{13}$~cm$^{-3}$ (essentially non-doped), $n_{3D} \sim 10^{18}$~cm$^{-3}$ (low-doped) 
and $n_{3D} \sim 10^{20}$~cm$^{-3}$ (highly-doped), determined by bulk-sensitive techniques (see Methods).  
Their large doping difference is already revealed by their photographs, shown in Figs.~2(a-c): 
while the non-doped sample is transparent, the highly-doped sample is shiny black.

Our first striking observation, shown in figure~2(d), is that even for the transparent non-doped sample [Fig.~2(a)],
for which no bulk bands at $E_F$ are expected, 
the cleaved surface yields intense strongly dispersive bands across the Fermi level. They correspond
to a large density of mobile carriers (see later) ensuring that, despite the electron emission, 
there is no charging of the surface of this otherwise highly insulating sample in the bulk. 
Equally stunning is the fact, shown in figures~2(d-f), that for all the studied bulk dopings
the electronic states around $E_F$, 
and in particular the bandwidths ($W = E_F -$~band~bottom) and Fermi momenta ($k_F$) of the observed bands, 
are essentially identical. 
All these observations contrast sharply with the large differences that would be expected 
from electron-doping in the bulk [Fig.~1(c)], where the 3D density of carriers scales as $k_{F}^{3}$.
%
The immediate conclusion is that the observed bands correspond to 2D states (see also the Supplementary Material). 

The spectra in figures~2(d-f), taken around $\Gamma_{102}$ and $\Gamma_{012}$ [B-like points in Fig.~2(j)], 
clearly show two strongly dispersing bands, termed `upper and lower parabolic bands'. 
Through a closer examination, we identify other two weakly dispersing bands, termed `upper and lower shallow bands'.
Their presence is revealed either by rotating the light polarization at right angles with respect to the measurement direction,
as illustrated for the non-doped sample in figure~2(g), or by measuring around a $\Gamma$-point 
at a larger emission angle in a different Brillouin zone [$\Gamma_{112}$, or point C in Fig.~2(j)],
as shown for the low-doped sample in figures~2(h-i). 
Note furthermore, from figures~2(d, g) that the changes in light polarization affect differently
the two strongly dispersing parabolic bands --completely suppressing the upper parabolic band
in the case of figure~2(g), while the increase in emission angle [compare Fig.~2(e) to Figs.~2(h, i)]
enhances in the same asymmetric way the negative-$k$ intensities of the two shallow bands and weakens the parabolic bands.
These dichroic effects, due to dipole-transition selection rules~\cite{Hufner-Book, Aiura-2002}, 
unambiguously show that the two parabolic bands have different symmetries,
which are also different from a third symmetry common to the two shallow subbands (see also the Supplementary Material).
Figure~3 illustrates the complete set of four bands overlaid on the data.

Let us now analyze quantitatively this band structure.
The upper and lower parabolic bands [Figs.~2(d-f)] have bandwidths of 210~meV and 100~meV, 
and Fermi momenta of $0.21$~\AA$^{-1}$ and $0.13$~\AA$^{^-1}$, respectively.
Thus, they correspond to light carriers with effective masses $m^{\star}_{y} \approx 0.7 m_e$ 
along $k_y$ ($m_e$ is the free electron mass).
The upper shallow band [Figs.~2(g, h)] has a bandwidth of 40~meV
and a Fermi momentum $k_F \approx 0.3 - 0.4$~\AA$^{-1}$, corresponding to heavy carriers with 
$m^{\star}_{y} \approx 10 - 20 m_e$.
The lower shallow band [Fig.~2(j)] disperses from about $-160$~meV at $\Gamma$ 
to about $-120$~meV at the Brillouin-zone boundary, and is thus fully occupied.
All the above figures differ by less than 10\% between the lowest- and highest-doping samples.
As shown in figure~4, the lower parabolic band generates a circular Fermi surface of radius $0.21$~\AA$^{-1}$ 
while the upper parabolic band and the upper shallow band form ellipsoids along $k_x$ and $k_y$, respectively, 
of semi axes $0.13$~\AA$^{^-1}$ and $0.3 - 0.4$~\AA$^{-1}$,
yielding a 2D 4-pronged Fermi surface.

From the area enclosed by each Fermi surface ($A_F$), the corresponding 2D carrier density is $n_{2D} = A_F/(2\pi^2)$.
This gives $7 \times 10^{13}$~cm$^{-2}$, or about 0.11 electrons per $a^2$, for the circular Fermi surface,
and $\sim 6.2 - 8.3 \times 10^{13}$~cm$^{-2}$, or about $0.1 - 0.13$ electrons per $a^2$, 
for each of the elliptical Fermi surfaces.
We then obtain $\sim 0.31 - 0.36$ electrons per $a^2$ (or about $2 \times 10^{14}$~cm$^{-2}$) 
coming from all the observed bands that cross $E_F$.

We now rationalize the observed electronic states as the subbands of a 2DEG
confined within a few unit cells at the surface of SrTiO$_3$. 
To this end, we consider a potential $V_0$ at the surface that confines the electronic motion along $z$ [inset of Fig.~1(d)]. 
This lowers the energy of the bands by about $V_0$, similar to the `band-bending' at semiconductor heterostructures, 
and produces an energy splitting between the different eigenstates that is inversely proportional 
to their effective masses along $z$ ($m^{\star}_{z}$). The resulting subband structure, depicted in Fig.~1(d), 
consists on a single $d_{xy}$-like band and two $d_{xz}/d_{yz}$ bands that are degenerate at $\Gamma$. 
As the $d_{xy}$ band has a very large $m^{\star}_{z}$, the attractive confining potential will merely pull it beneath $E_F$ 
(its energy-splitted eigenstates will have a negligible separation).
Thus, we identify this band with the lower parabolic band in our spectra of figure~2, 
and note it $E_1 (d_{xy})$. On the other hand, the $d_{xz}$ and $d_{yz}$ subbands, light along $z$, 
will show large energy splittings. They are noted $E_n (d_{xz/yz})$ ($ n = 1, 2, \dots$) in figure~1(d). 

Other effects beyond this simplified analysis, such as spin-orbit coupling 
and/or the low-temperature tetragonal~\cite{Lytle-1964, Mattheiss-1972-CT-Transition} 
and possibly orthorhombic~\cite{Lytle-1964} distortions, 
can lift the degeneracy between the $d_{xz}$ and $d_{yz}$ subbands, as represented in Fig.1~(e). 
This would induce a coupling, resulting in hybridization 
between the light and heavy subbands, as indeed evidenced by our data for the lower parabolic band and the shallow bands 
(Fig.~3, LV polarization). 
Another possibility would be a surface reconstruction, although this is not evidenced by our data 
--which follows the periodicity of the unreconstructed bulk lattice without apparent band folding.
Resolving the detailed origin of such degeneracy lift, which is not relevant to the 
discoveries reported here, will require future investigations.
Hence, we identify the lower shallow band and upper parabolic band in our spectra 
(Fig.~2) with the splitted doublet $E_1 (d_{xz})$ and $E_1 (d_{yz})$, respectively. 
From the data, the doublet splitting is $\Delta \approx 60$~meV. We finally identify the upper shallow band with $E_2 (d_{xz})$. 
Given that the bottom of this band is at $-40$~meV, its upper partner $E_2 (d_{yz})$ 
would occur at about $+20$~meV, and cannot be observed.

This subband hierarchy respects the symmetry considerations previously deduced
from the effects of light polarization and emission angle, rendering additional support to the picture of a confined 2DEG.

To characterize quantitatively this 2DEG, we use a simple schematic model. 
We assume that the surface has a homogeneous positive charge density, generating a confining electric field $F$ 
inside the solid, and thus a triangular wedge potential of depth $V_0$ at the surface ($z = 0$)
and value $V(z) = -V_0 + eFz$ along $z$, as depicted in the inset of Fig.~1(d). 
The details of the calculations are presented in the Supplementary Information. 
From the experimental splitting $E_2(d_{xz}) - E_1(d_{xz}) = 120$~meV
between the first two subbands of the $d_{xz}$ orbitals, 
and their light mass $m^{\star}_{z} \approx 0.7 m_e$ along $z$, we deduce a strong confining field $F \approx 83$~MV/m. 
From $E_1(d_{xy})$, the confining potential is estimated to be $V_0 \approx -260$~meV. 
The deduced values of $V_0$ and $F$, and the measured effective masses and  
doublet splitting of 60 meV between the $d_{xz}$ and $d_{yz}$ subbands, suffice to retrieve
all the observed subband energies.
The width $L$ of the confined 2DEG results from the spatial extension of the highest occupied state $E_2(d_{xz})$,
or independently from the approximate expression $eFL \approx E_2(d_{xz}) - E_1(d_{xz})$ [inset of Fig.~1(d)],  
yielding consistent values in the range $L \approx 14.5 - 18$~\AA~$\approx 4 - 5$~ unit cells. 
Additionally, from the confining field $F$ and the polarizability of the medium, we calculate an induced 
surface charge density of $\sim 0.25$ electrons per $a^2$, in good agreement with the experimental value. 

The band-bending due to confinement should induce an energy down-shift of about $V_0$ of the O-$2p$ valence band, 
creating a gap with respect to $E_F$ that is larger than the optical gap. 
This effect has been reported in previous experiments~\cite{Aiura-2002, Takizawa-O2pBands-2009}, 
and is also observed in our data (Supplementary Material). 
Aiura and coworkers~\cite{Aiura-2002} demonstrated that it is due to surface oxygen-vacancies, 
which appear thus as the most reasonable candidate to explain the origin of the 2DEG observed in our experiments. 
In fact, cleaving is likely to produce a massive removal of surface oxygen, much larger than and basically independent of 
the bulk concentration of vacancies, providing 2 dopant electrons per created vacancy. 
These two electrons will delocalize within the potential wedge created by the positively charged layer 
of surface vacancies. Note that cleaving this almost cubic oxide system necessitates overcoming 
the strong binding electric forces between atoms, equivalent to inducing a `mechanical dielectric breakdown'. 
Interestingly, from our data analysis, the large electric field generated at the surface of SrTiO$_3$ 
is of the order of typical dielectric breakdown fields~\cite{Barret-Bkdwn-STO-1964}, 
and thus compatible with a cleaved surface in this material.

To give additional support to our picture and data analyses, 
we performed {\it ab-initio} calculations including oxygen vacancies at the surface. The results are
consistent with the above simplified wedge-potential model (Supplementary Material).

Strikingly, the sheet carrier density, the confinement size, the confining field $F$ and potential $V_0$, 
and the presence of both mobile and quasi-localized carriers in this novel electron gas 
at the surface of SrTiO$_3$ compare well with the characteristics of other 2DEG's 
at different types of SrTiO$_3$ interfaces~\cite{Thiel-TunableQ2DEG-Science-2006, Reyren-SC-interface-LAO-STO-Science-2007,
Ueno-Electric-Field-Induced-SC-Plastic-STO-2008, Popovic-LAO-STO-FS-ElectronicStructure-PRL2008, 
Thales-CAFM-Size2DEG-2008, Copie-PRL-2009, Hwang-Multiple-Carriers-LAO-STO-2009, Claessen-HAXPES-LAO-STO-2009}. 
All these remarkable consistencies suggest that all these 2DEG's may be understood on a common basis. 
In particular, note that different electron confinement mechanisms, such as the electric field created by a polar 
LaAlO$_3$ layer~\cite{Thiel-TunableQ2DEG-Science-2006}
or the direct application of an electric field~\cite{Ueno-Electric-Field-Induced-SC-Plastic-STO-2008},
would recreate a 2DEG confined within a nanometric layer of SrTiO$_3$ with a subband electronic structure 
similar in nature to the one directly unveiled by our measurements.

Our experiments directly show that the Ti-$3d_{xy}$ states of light and highly mobile carriers 
become the first available levels for the 2DEG at the surface of SrTiO$_3$. 
Altogether, our results demonstrate 
the fundamental paradigm for understanding the properties of oxide surfaces and heterostructures,
namely the `electronic reconstruction' into several partially-occupied subbands 
at the near-interface region~\cite{Okamoto-Millis-2004, Popovic-LAO-STO-FS-ElectronicStructure-PRL2008, 
Pentcheva-LAO-STO-ElectronicReconstruction-2008, Ueno-Electric-Field-Induced-SC-Plastic-STO-2008,
Salluzzo-XAS-OrbitalReconstruction-2009}.
Our discoveries also raise the appealing possibility that oxygen vacancies may be induced to aggregate 
or self-organize at the surface of SrTiO$_3$ or related interfaces -a situation that might be at the origin 
of the reversible patterning of conducting nano-lines in LaAlO$_3$/SrTiO$_3$ 
interfaces~\cite{Manhart-OxideNanoElectronics-2009, Cen-NanoPatterning-LAO-STO-2008}. 
In a similar way, nano-patterning of oxygen vacancies at the surface of SrTiO$_3$ could provide a model system 
for low dimensional confined electron gases. More generally, we suggest that the engineering 
of atom-vacancies at the surfaces of transition metal oxides will open a new avenue 
in correlated-electron surface science.

\section{methods}

\subsection{Sample preparation and measurement technique}
%
High-quality single crystals of SrTiO$_3$ were doped with oxygen vacancies by high-temperature treatment 
in low-oxygen pressure, except for the non-doped sample, which was a bare SrTiO3 substrate 
measured without any previous preparation. 
The bulk carrier concentrations were deduce from Hall and resistivity measurements.
 
The ARPES experiments were done at the Synchrotron Radiation Center (SRC, University of Wisconsin, Madison)
and at the Synchrotron SOLEIL (France), 
using 47~eV (SRC) and 45~eV (SOLEIL) linearly polarized photons, 
a Scienta 2002 detector with horizontal slits for the highly-doped sample (SRC), 
and a Scienta R4000 with vertical slits for the low-doped (SRC) and non-doped (SOLEIL) samples. 
The above photon energies, close to resonant photoemission at the Ti $3p \rightarrow 3d$ edge, 
yield intense quasi-particle peaks in SrTiO$_3$~\cite{Aiura-2002}.
%
The momentum and energy resolutions were $0.25^{\circ}$ and 20~meV, respectively. 
The diameter of the incident photon beam was smaller than $100$~$\mu$m. 
The samples were cleaved {\it in-situ} along the $c$-axis at 20~K (SRC) and 10~K (Soleil), 
in pressure lower than $6\times10^{-11}$~Torr. 
%
For each set of measurements, one narrow highly-emitting terrace was kept fixed.
The results have been reproduced in at least two cleaves for each bulk doping.

\subsection{Measurement geometries}
The sample's surface defines the $xy$-plane. 
The measurements 
on the $k$-space hemisphere 
were taken along $(010)$ (or $k_y$). 
The data in Figs.~2(d, g) and Fig.~3 were collected in B-like points slightly above $\Gamma_{102}$, 
with the photon polarization along $y$ (LV) in Fig.~2(d), and along the $xz$-plane (LH) in Fig.~2(h). 
The data in Figs.~2(e, f) [Figs.~2(h, i)] correspond to B-like (C-like) points slightly above $\Gamma_{012}$ 
(slightly below $\Gamma_{112}$). For these points, the photon polarization is not parallel to any of the sample's 
symmetry directions or planes, and the spectra have different symmetry mixtures.
The data in Fig.~4 were collected with LH photon polarization. 
The Supplementary Information discusses further the measurement geometries and their corresponding dipole selection rules.


\newpage
\section{Acknowledgements}
We are grateful to Ralph Claessen, Yannick Fagot, Marc Gabay, Ingrid C. Infante, Daniel Malterre, 
Friedrich Reinert and Nicolas Reyren for discussions, and to Rub\'en Guerrero for help
with the transport measurements.
This work was supported by the ANR OXITRONICS (AFSS, OC, MB, MJR, AB).
The Synchrotron Radiation Center, University of Wisconsin-Madison, is supported by the National Science Foundation 
under award no. DMR-0537588.
RW is support by CONICET (grant PIP 112-200801-00047) and ANCTyP (grant PICT 837/07).

\section{Competing interests}  
The authors declare that they have no competing financial interests.

\section{Additional information}
Correspondence and requests for materials should be addressed to A.F.S.S.\\
 (e-mail:~andres.santander@csnsm.in2p3.fr).

\newpage
\begin{figure}
  \begin{center}
     \includegraphics[width=14cm]{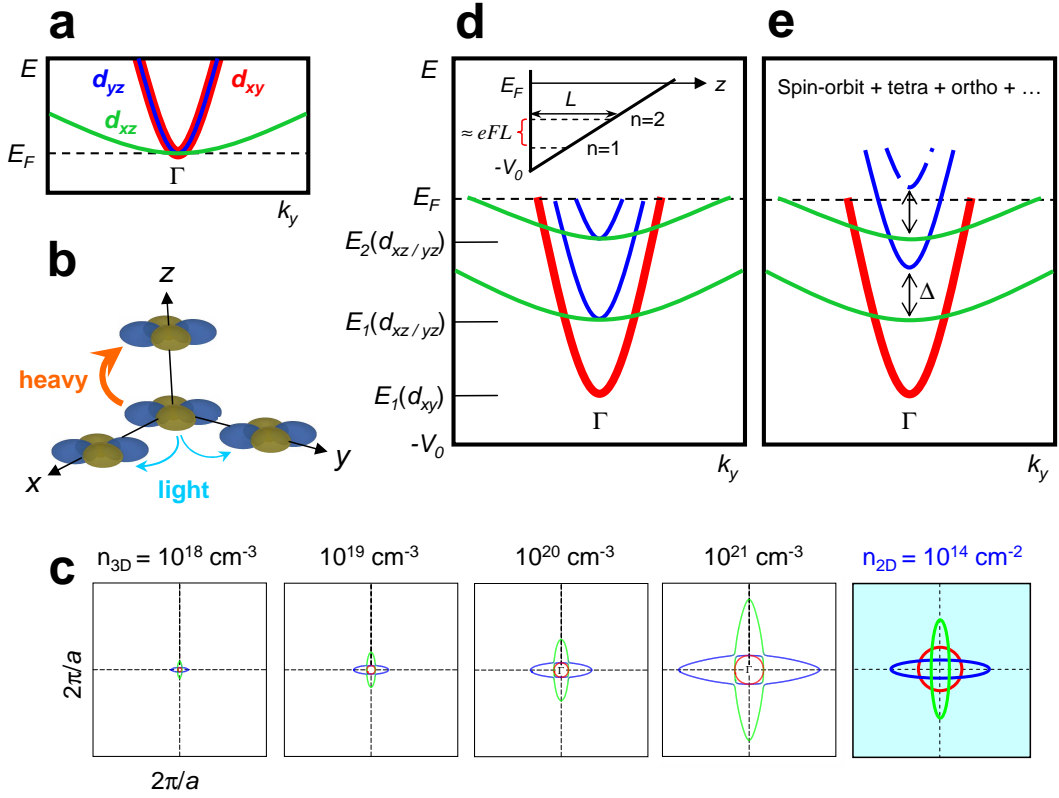}
  \end{center}
  \caption{\label{Fig1}
  		  \footnotesize{
		   	{\bf Schematic electronic structure of SrTiO$_3$ and effects of electron confinement}.
		    {\bf a},~Bulk conduction band of SrTiO$_3$ along $k_y$, consisting on a heavy $d_{xz}$ band (green)
		    and a doublet of light $d_{xy}-d_{yz}$ bands (red and blue). These bands stem from the small and large overlaps
		    of the Ti-$3d$ orbitals along the $y$ direction, as depicted in {\bf b} for the case of $3d_{xy}$ orbitals.
		    {\bf c},~Resulting 3D Fermi surfaces (cut along the $xy$-plane) for several bulk dopings.  
		    The last panel shows, for comparison, the 2D Fermi surface 
		    for a 2DEG of density $\sim 10^{14}$~cm$^{-2}$ 
		    (hybridization between different Fermi-surface sheets is not included), 
		    deduced from a tight-binding model~\cite{Popovic-LAO-STO-FS-ElectronicStructure-PRL2008}.
		    Colors indicate the character of each Fermi surface sheet along $k_y$. 
		    {\bf d},~Quantum-well states, or subbands, resulting from the confinement of electrons near the surface
		    of SrTiO$_3$. 
		    The inset shows a wedge-like potential created by an electric field of strength $F$ at the surface, 
		    used as a simple model to analyse the ARPES data (main text). The size $L$ of the confined 2DEG
		    can be estimated from the extension of the highest occupied state ($n=2$ in our case) or approximately from 
		    the energy difference between the lowest ($n=1$) and highest occupied states. 
		    {\bf e},~Additional degeneracy lifts at $\Gamma$ occur due to spin-orbit coupling,
		    tetragonal and orthogonal distortions, or possible surface reconstructions. 
		   }
		  }
\end{figure}

\begin{figure}
  \begin{center}
     \includegraphics[width=14cm]{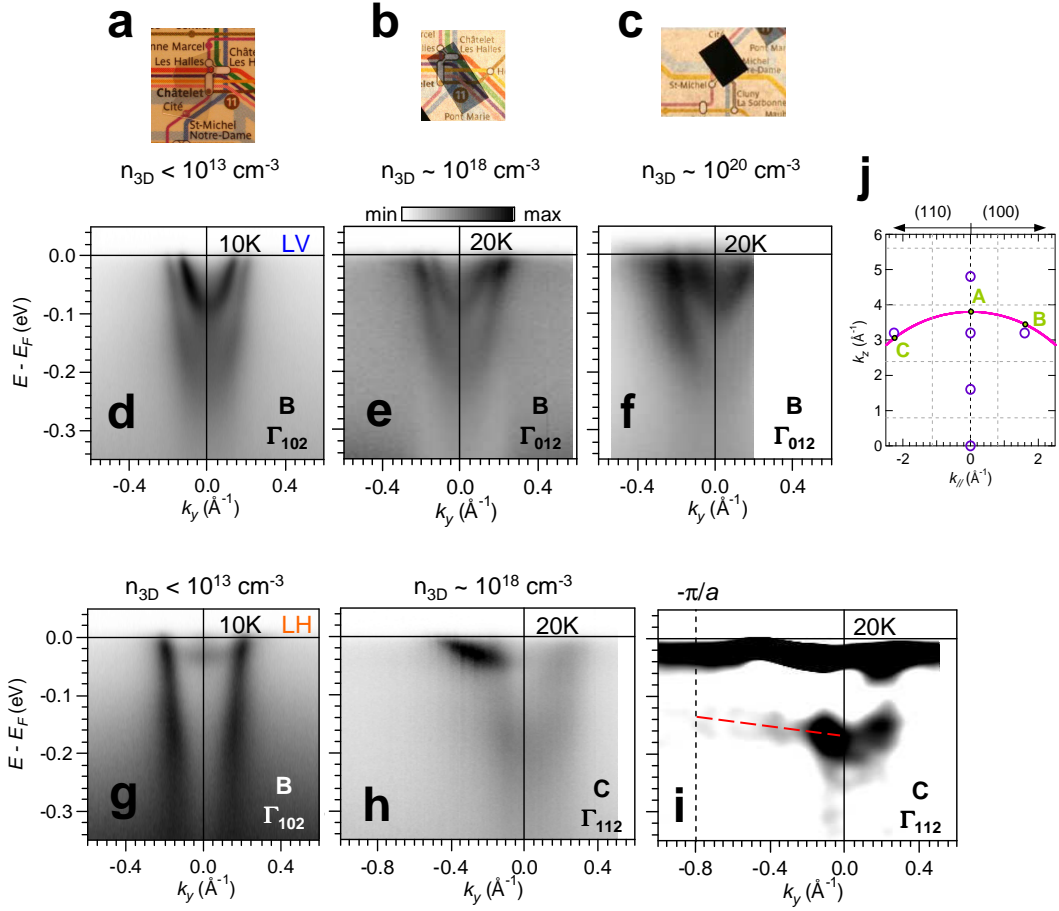}
  \end{center}
  \caption{\label{Fig2}
		  \footnotesize{
  		   {\bf Universal electronic structure at the surface of SrTiO$_3$}.
		   {\bf a-c},~Photographs of the samples studied by ARPES. Their bulk dopings are written beneath.
		   {\bf d-f},~Corresponding energy-momentum intensity maps close to the $\Gamma$ point
		   (B-like points in panel {\bf j}).
		   {\bf g},~Effects of changing the photon polarization (specifications below) from linear vertical 
		   (LV, panel {\bf d}) to linear horizontal (LH), illustrated here for the non-doped sample. 
		   The upper shallow band appears, and the upper parabolic band disappears. 
		   The effects of collecting electrons around a $\Gamma$-point at a larger emission angle
		   in a different Brillouin zone (point C in panel {\bf j}) are also shown, 
		   for the case of the low-doped sample, in the energy-momentum intensity map of panel {\bf h},
		   and on its second derivative (only negative values shown) in panel {\bf i}, where the dispersion
		   of the lower shallow band is indicated by a dashed red line. 
		   Here, the negative-$k$ intensities of the shallow bands are enhanced, while the intensities of the
		   parabolic bands decrease.
		   {\bf j},~2D representation of the 3D Brillouin zones of SrTiO$_3$ (dashed lines) in relation to the hemispherical
		   measurement surface in momentum-space at $h\nu = 47$~eV~\cite{Aiura-2002, Shin-SoftX-ARPES-STO-IGS}. 
		   The measurement temperature (10~K or 20~K) is indicated in each panel. 
		   }		   
		   }
\end{figure}

\begin{figure}
  \begin{center}
     \includegraphics[width=14cm]{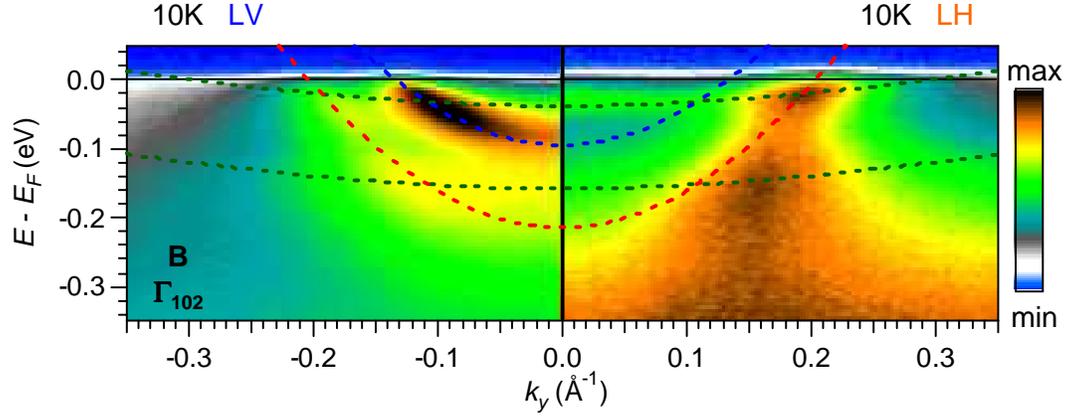}
  \end{center}
  \caption{\label{Fig3}
  		  \footnotesize{
		   {\bf Summary of subbands for the 2DEG at the surface of SrTiO$_3$}.
		   Side-by-side representation of the LV and LH spectra
		   around $\Gamma_{102}$ for the non-doped sample. 
		   An elongated aspect ratio has been used to better discern the shallow bands.
		   The dotted lines are tight-binding representations
		   of the bands~\cite{Popovic-LAO-STO-FS-ElectronicStructure-PRL2008}, following the same color
		   scheme as in Figs.~1(d,~e).  
		   Hybridizations between the lower parabolic band and the shallow bands are observed on the spectra
		   taken with LV polarization.
		   }
		   }
\end{figure}

\begin{figure}
  \begin{center}
     \includegraphics[width=14cm]{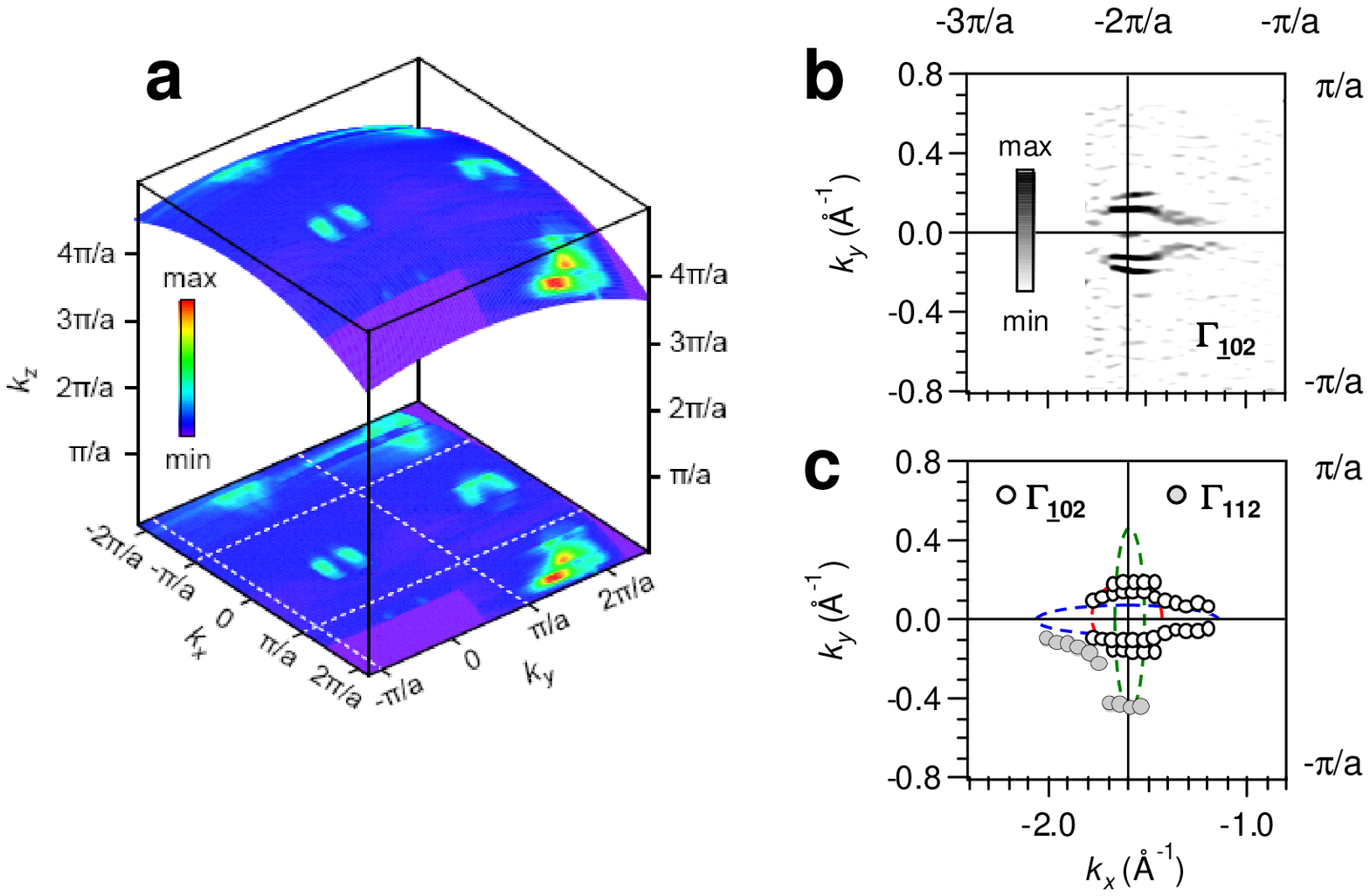}
  \end{center}
  \caption{\label{Fig4}
		  \footnotesize{ 
  		   {\bf Fermi surface of 2DEG at the surface of SrTiO$_3$}.
		   {\bf a},~ARPES Fermi surface map (integrated within $\pm 20$~meV around $E_F$) across several Brillouin
		   zones for a photon energy $h\nu = 47$~eV, displayed on the spherical shell of 3D $k$-space
		   and projected on the $k_{x}-k_{y}$ plane. The Fermi-surface circle from the lower ($d_{xy}$-like) 
		   light parabolic band is best observed in the $\Gamma_{002}$ and $\Gamma_{012}$ (and equivalent) Brillouin zones, 
		   while the two elliptical perpendicular Fermi surfaces from the upper parabolic band ($d_{yz}$) 
		   and the upper shallow band ($d_{xz}$) are best observed off-normal emission 
		   ($\Gamma_{012}$, $\Gamma_{112}$ and equivalent). 
		   {\bf b},~Second derivative (negative values only) of the Fermi surface map over $\Gamma_{\underline{1}02}$,
		   showing portions of the circular and the horizontal elliptical Fermi surfaces. The intensity of the vertical
		   ellipse is very weak here due to dipole selection rules.  
		   {\bf c},~Collected Fermi momenta from high-statistics measurements around $\Gamma_{\underline{1}02}$ (white circles)
		   and $\Gamma_{112}$ (grey circles), and comparison to the Fermi surfaces resulting from the tight-binding
		   bands of figure~3 (colors indicate the character along $k_y$ and follow the scheme of precedent figures). 
		   All panels correspond to measurements on the low-doped sample.
		   }
		   }
\end{figure}

\begin{references}
\bibitem{Ramirez-NV-QHE-ZnO-2007} Ramirez, A. P. Oxide electronics emerge. {\it Science} {\bf 315}, 1377 (2007).
\bibitem{Manhart-OxideNanoElectronics-2009} Cen, C., Thiel, S., Mannhart, J. \& Levy, J..
		Oxide nanoelectronics on demand. {\it Science} {\bf 323}, 1026 (2009).
\bibitem{Ohtomo-Wang-2004} Ohtomo, A. \& Hwang, H. Y. A high-mobility electron gas 
		at the LaAlO$_3 /$SrTiO$_3$ heterointerface. {\it Nature} {\bf 427}, 423 (2004).
\bibitem{Ahn-2003} Ahn, C. H., Triscone, J.-M  \&  Mannhart, J. Electric field effect in correlated oxide systems.
		{\it Nature} {\bf 424}, 1015 (2003).  
\bibitem{Ohtomo-2002} Ohtomo, A., Muller, D. A., Grazul, J. L. \&  Hwang, H. Y. 
		Artificial charge-modulation in atomic-scale perovskite titanate superlattices. 
		{\it Nature} {\bf 419}, 378 (2002).
\bibitem{Thiel-TunableQ2DEG-Science-2006} Thiel, S. , Hammer, G., Schmehl, A., Schneider, C. W. \& Mannhart, J.
		Tunable quasi–two-dimensional electron gases in oxide heterostructures.
		{\it Science}~{\bf 313}, 1942-1945 (2006). 
\bibitem{Cen-NanoPatterning-LAO-STO-2008} Cen, C. {\it et al.} Nanoscale control of an interfacial
		metal–insulator transition at room temperature. {\it Nature Mater.} {\bf 7}, 298-302 (2008).
\bibitem{Reyren-SC-interface-LAO-STO-Science-2007} Reyren, N. {\it et al.} Superconducting interfaces 
		between insulating oxides. {\it Science}~{\bf 317}, 1196-1199 (2006).
\bibitem{Ueno-Electric-Field-Induced-SC-Plastic-STO-2008} Ueno, K. {\it et al.}
		Electric-field-induced superconductivity in an insulator.
		{\it Nature Mater.} {\bf 7}, 855-858 (2008).
\bibitem{Brinkman-MagneticInterface-LAO-STO-NatMat-2007} Brinkman, A. {\it et al.}
		Magnetic effects at the interface between non-magnetic oxides.
		{\it Nature Mater.}~{\bf 6}, 493-496 (2007).
\bibitem{Okamoto-Millis-2004} Okamoto, S. \& Millis, A. J. 
		Electronic reconstruction at an interface between a Mott insulator and a band insulator.
		{\it Nature} {\bf 428}, 630 (2004).
\bibitem{Popovic-LAO-STO-FS-ElectronicStructure-PRL2008} Popovi\'c, Z. S., Satpathy, S. \& Martin, R. M. 
		Origin of the Two-Dimensional Electron Gas Carrier Density at the LaAlO$_3$ on SrTiO$_3$ Interface.
		{\it Phys. Rev. Lett} {\bf 101}, 256801 (2008).	
\bibitem{Thales-CAFM-Size2DEG-2008} Basletic, M. {\it et al.} Mapping the spatial distribution of charge
		carriers in LaAlO$_3 /$SrTiO$_3$ heterostructures. {\it Nature Mater.} {\bf 7}, 621-625 (2008).
\bibitem{Tokura-Parylene-FET-2009} Nakamura, H. {\it et al.} Tuning of metal-insulator transition of 
		two-dimensional electrons at parylene/SrTiO$_3$ interface by electric field.
		{\it J.~Phys.~Soc.~Jpn.}~{\bf 78}, 083713 (2009). 
\bibitem{Mattheiss-1972-STO} Mattheiss, L. F. Energy bands for KNiF$_3$, SrTiO$_3$, KMoO$_3$ and KTaO$_3$.
		{\it Phys.~Rev.~B}~{\bf 6}, 4718 (1972).
\bibitem{Hufner-Book} H\"ufner, S. Photoelectron spectroscopy: principles and applications.
		Third edition, Springer (2003).
\bibitem{Fujimori-1992-LaSrTiO3} Fujimori, A. {\it et al.} Doping-induced changes in the electronic structure 
		of La$_x$Sr$_{1-x}$TiO$_3$: limitation of the one-electron rigid-band model and the Hubbard model.
		{\it Phys. Rev. B} {\bf 46}, 9841 (1992).
\bibitem{Aiura-2002} Aiura, Y. {\it et al.} Photoemission study of the metallic state of lightly electron-doped SrTiO$_3$. 
		{\it Surf. Sci.} {\bf 515}, 61 (2002).		
\bibitem{Shin-SoftX-ARPES-STO-IGS} Ishida, Y. {\it et al}. Coherent and Incoherent Excitations of Electron-Doped SrTiO$_3$.
		{\it Phys.~Rev.~Lett.}~{\bf 100}, 056401 (2008).
\bibitem{Takizawa-O2pBands-2009} Takizawa, M. {\it et al.} Angle-resolved photoemission study of Nb-doped SrTiO$_3$. 
		{\it Phys. Rev. B} {\bf 79}, 113103 (2009).
\bibitem{Meevasana-ARPES-STO-2010} Meevasana, W. {\it et al.} Strong energy-momentum dispersion 
		of phonon dressed carriers in the lightly doped band insulator SrTiO$_3$.
		{\it New~J.~Phys.}~{\bf 12}, 023004 (2010).
%
\bibitem{Rotenberg-ARPES-STO-CubicTetra-2010} Chang, Y. J., Bostwick, A., Kim, Y. S., Horn, K. \& Rotenberg, E.
		Structure and correlation effects in semiconducting SrTiO$_3$.
		{\it Phys.~Rev.~B}~{\bf 81}, 235109 (2010).
\bibitem{Lytle-1964} Lytle, F. W. X-ray diffractometry of low-temperature phase transformations 
		in strontium titanate. {\it J.~Appl.~Phys.}~{\bf 35}, 2212-2215 (1964).
\bibitem{Mattheiss-1972-CT-Transition} Mattheiss, L. F. Effect of the 110$^\circ$~K transition on the SrTiO$_3$
		conduction bands. {\it Phys.~Rev.~B}~{\bf 6}, 4740 (1972).
%
\bibitem{Barret-Bkdwn-STO-1964} Barret, H.~H. Dielectric breakdown of single-crystal strontium titanate.
		{\it J.~Appl.~Phys.}~{\bf 35}, 1420-1425 (1964).
\bibitem{Copie-PRL-2009} Copie, O. {\it et al.} Towards two-dimensional metallic behavior 
		at LaAlO$_3 /$SrTiO$_3$ interfaces. {\it Phys.~Rev.~Lett.}~{\bf 102}, 216804 (2009).	
\bibitem{Hwang-Multiple-Carriers-LAO-STO-2009} Seo, S. S. A. {\it et al.}
		Multiple conducting carriers generated in LaAlO$_3/$SrTiO$_3$ heterostructures.
		{\it Appl.~Phys.~Lett.}~{\bf 95}, 082107-082109 (2009).
%
\bibitem{Claessen-HAXPES-LAO-STO-2009} Sing, M. {\it et al.} 
		Profiling the interface electron gas of LaAlO$_3/$SrTiO$_3$ heterostructures 
		with hard X-ray photoelectron spectroscopy. 
		{\it Phys.~Rev.~Lett.}~{\bf 102}, 176805 (2009).
\bibitem{Pentcheva-LAO-STO-ElectronicReconstruction-2008} Pentcheva, R. \& Pickett, W. E. 
		Ionic relaxation contribution to the electronic reconstruction at the n-type LaAlO$_3/$SrTiO$_3$ interface.
		{\it Phys. Rev. B} {\bf 78}, 205106 (2008).
\bibitem{Salluzzo-XAS-OrbitalReconstruction-2009} Salluzzo, M. {\it et al.} Orbital reconstruction and the 
		two-dimensional electron gas at the LaAlO$_3/$SrTiO$_3$ interface.
		{\it Phys.~Rev.~Lett.}~{\bf 102}, 166804 (2009). 	
\end{references}
\end{document}


\title{SUPPLEMENTARY MATERIAL\\
       	Unveiling a two-dimensional electron gas with universal subbands at the surface of SrTiO$_3$
	  }
\author{A.~F.~Santander-Syro$^{1,2,\star}$, 
		O.~Copie$^{3,4}$,
		T.~Kondo$^{5}$, 
		F.~Fortuna$^{1}$,
		S.~Pailh\`es$^{6}$,
		R.~Weht$^{7,8}$,
		X.~G.~Qiu$^{9}$,
		F.~Bertran$^{10}$,
		A.~Nicolaou$^{10}$,
		A. Taleb-Ibrahimi$^{10}$,
		P. Le F\`evre$^{10}$,
		G.~Herranz$^{11}$,
		M.~Bibes$^{3}$,
		Y.~Apertet$^{12}$,
		P.~Lecoeur$^{12}$,
		M.~J.~Rozenberg$^{13}$
		\& A.~Barth\'el\'emy$^{3}$. 
		}

\address{\footnotesize{
		 $^1$CSNSM, Universit\'e Paris-Sud and CNRS/IN2P3
		 B\^atiments 104 et 108, 91405 Orsay cedex, France.
 	   	 \\
		 $^2$Laboratoire Photons Et Mati\`ere, UPR-5 CNRS, ESPCI,
 	   	 10 rue Vauquelin, 75231 Paris cedex 5, France.
		 \\
		 $^3$Unit\'e Mixte de Physique CNRS/Thales, Campus de l'Ecole Polytechnique, 1 Av. A. Fresnel, 
		 91767 Palaiseau, France and Universit\'e Paris-Sud, 91405 Orsay, France.
		 \\
		 $^4$Universit\"at W\"urzburg, Experimentelle Physik VII, 97074 W\"urzburg, Germany.
		 \\
		 $^5$Ames Laboratory and Department of Physics and Astronomy, Iowa State University,
		 Ames, IA 50011.
		 \\
		 $^6$Laboratoire L\'eon Brillouin, CEA-CNRS, CEA-Saclay, 91191 Gif-sur-Yvette, France.
		 \\
		 $^7$Gerencia de Investigaci\'on y Aplicaciones, CNEA, 1650 San Mart\'in, Argentina.
		 \\
		 $^8$Instituto S\'abato, Universidad Nacional de San Mart\'in, CNEA, 1650 San Mart\'in, Argentina. 
		 \\
		 $^9$Institute  of Physics and National Laboratory for Condensed Matter Physics,
		 Chinese Academy of Sciences, Zhongguancun nansanjie 8, Beijing 100190, China.
		 \\
		 $^{10}$Synchrotron SOLEIL, L'Orme des Merisiers, Saint-Aubin-BP48, 91192 Gif-sur-Yvette, France.
		 \\
		 $^{11}$Institut de Ci\`encia de Materials de Barcelona, CSIC, Campus de la UAB, 
		 08193 Bellaterra, Catalunya, Spain.
		 \\
		 $^{12}$Institut d'Electronique Fondamentale, Universit\'e Paris-Sud, B\^atiment 220, 91405 Orsay, France.
		 \\
		 $^{13}$Laboratoire de Physique des Solides, Universit\'e Paris-Sud, B\^atiment 510, 91405 Orsay, France.
		 \\
		 $^{\star}$e-mail:~andres.santander@csnsm.in2p3.fr  
		}
		}
\begin{abstract}
	The angle-resolved photoemission spectroscopy (ARPES) data on SrTiO$_3$ 
	show the existence of a two-dimensional electron gas (2DEG) beneath the surface of this material.  
	We provide here additional evidence of its 2D-like nature,
	showing that the bands do not disperse along the $k_z$ [or $(001)$] direction,
	although their intensity is strongly modulated as a function of photon energy, photon polarization 
	and emission angle for different Brillouin zones. 
	%
	%
	We also provide the details of the modeling of these bands in terms of a confining wedge-like potential well. 
	Additionally, we show that in our data the O-$2p$ valence band is shifted to larger binding energies
	with respect to what is expected from the bulk band gap (`band bending'), 
	by an amount comparable to the confining potential depth.
	We finally present {\it ab-initio} calculations that support the picture of the formation 
	of a 2DEG due to oxygen vacancies at the surface of SrTiO$_3$.  
\end{abstract}
\maketitle
           
\subsection{Photon energy dependence of the electronic subbands at the surface of SrTiO$_3$}

\begin{figure}
    \begin{center}
    	\includegraphics[width=12cm]{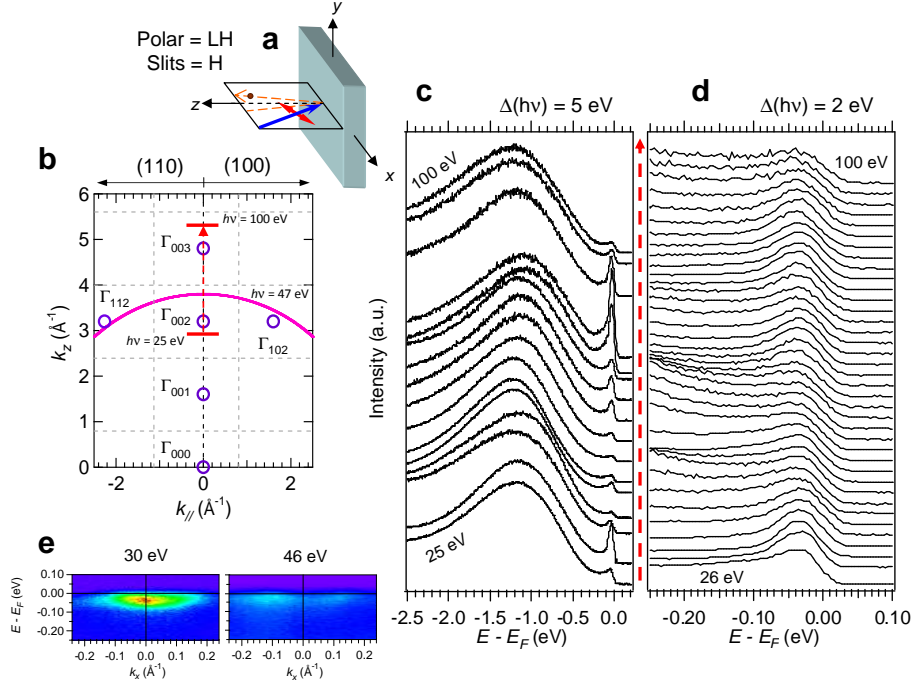}
    \end{center}
    \caption{\label{hv-dependence}
    		\footnotesize{
            {\bf Photon-energy dependence of the ARPES spectra of SrTiO$_3$.}
            {\bf a},~Schematics of the experimental geometry for the data in this figure.
            The sample is the dark parallelepiped, the $xz$ mirror plane is transparent,
            the light beam is represented by the blue arrow, its polarization by the red double arrow,
            and the collected electron beam by the dashed orange arrows. The same conventions are used
            in the rest of the figures in this Supplement.
            {\bf b},~Representation of the 3D reciprocal space. The region covered by varying
            the photon energy from 25~eV to 100~eV is shown red dashed arrow and the thick red lines. 
            The open circles are $\Gamma$ points, the dashed lines
            are Brillouin-zone edges, and the violet circle is the spherical shell on which the
            measurements at $h\nu = 47$~eV are carried out. To convert the photon energy dependence into
    		$k_z$ dependence, an `inner potential' of 12~eV was used, as deduced in previous works~\cite{Aiura-2002}.
            {\bf c},~Integrated photoemission spectra at normal emission over a wide energy range, 
            showing the oxygen-vacancy related intra-gap states at $1.3$~eV~\cite{Aiura-2002} 
            and the quasi-particle (QP) peak near $E_F$. Notice the large variations of the QP
            peak intensity as a function of photon energy (varying from 25~eV to 100~eV in steps of 5~eV).
            {\bf d},~Zoom over the QP peak in {\bf c}, after normalizing by the area of the peak.
            The peak does not disperse with photon energy ($26 - 100$~eV in steps of 2~eV), indicating its 2D character.  
            {\bf e},~Examples of two energy-momentum intensity maps at $h\nu = 30$~eV, at $\Gamma_{002}$,
            and $h\nu = 46$~eV, close to the zone edge. The band corresponds to the $d_{xz}$ states.
            Its intensity is almost suppressed at $h\nu = 46$~eV, but the dispersion is unchanged.
            }
            }
\end{figure}

Figure~SF~\ref{hv-dependence} presents data obtained at normal emission as a function of photon energy.
The data were collected using linear horizontal (LH) light polarization (polarization vector in the $xz$ plane),
and detection along $x$ [horizontal (H) slits] in the sample's $xz$-plane, which in this geometry also coincides
with the light incidence plane. 
The final electron state has to be even with respect to the sample's mirror plane 
to be detected by the analyzer (the $xz$-plane in this case) 
--otherwise the electron wavefunction would be zero in this plane. 
Hence, the dipole-transition selection rules imply that the initial state and the dipole operator 
(equivalently, the polarization vector) must have the same symmetry 
with respect to the mirror plane~\cite{Hufner-Book, Aiura-2002}.  
In the geometry of figure~SF~\ref{hv-dependence}, the light polarization is even under reflection on the $xz$ plane.
Thus, only initial states of even parity with respect to this plane can be detected. 
Hence, only the upper $d_{xz}$-like band, which is furthermore {\it light} along $x$, is observed in this geometry
(experimentally, the lower $d_{xz}$-like band has a very weak intensity in this geometry).

As observed in Fig.~SF~\ref{hv-dependence}(c, d), the upper $d_{xz}$-like band 
does not disperse as a function of photon energy, or equivalently as a function of $k_z$,
confirming its 2D surface-like character.
However, it presents a marked intensity modulation as a function of photon energy, 
a behaviour also observed in surface states and quantum well states
of simple metallic systems~\cite{Petroff-hv-dependence-SurfStates-1980, Chiang-hv-dependence-QWS-1999}. 
Thus, for instance, while the band is clearly seen at $h\nu = 30$~eV, its intensity is very weak
around $h\nu = 47$~eV, where the rest of our experiments were performed.
As explained in the main text, and also discussed below, at this photon energy the $d_{xz}$-like bands 
and the other subbands are distinctly observed in other Brillouin zones and other polarization/slits geometries.

\subsection{Intensity modulations of the electronic subbands with emission angle and photon polarization}

\begin{figure}
    \begin{center}
    	\includegraphics[width=12cm]{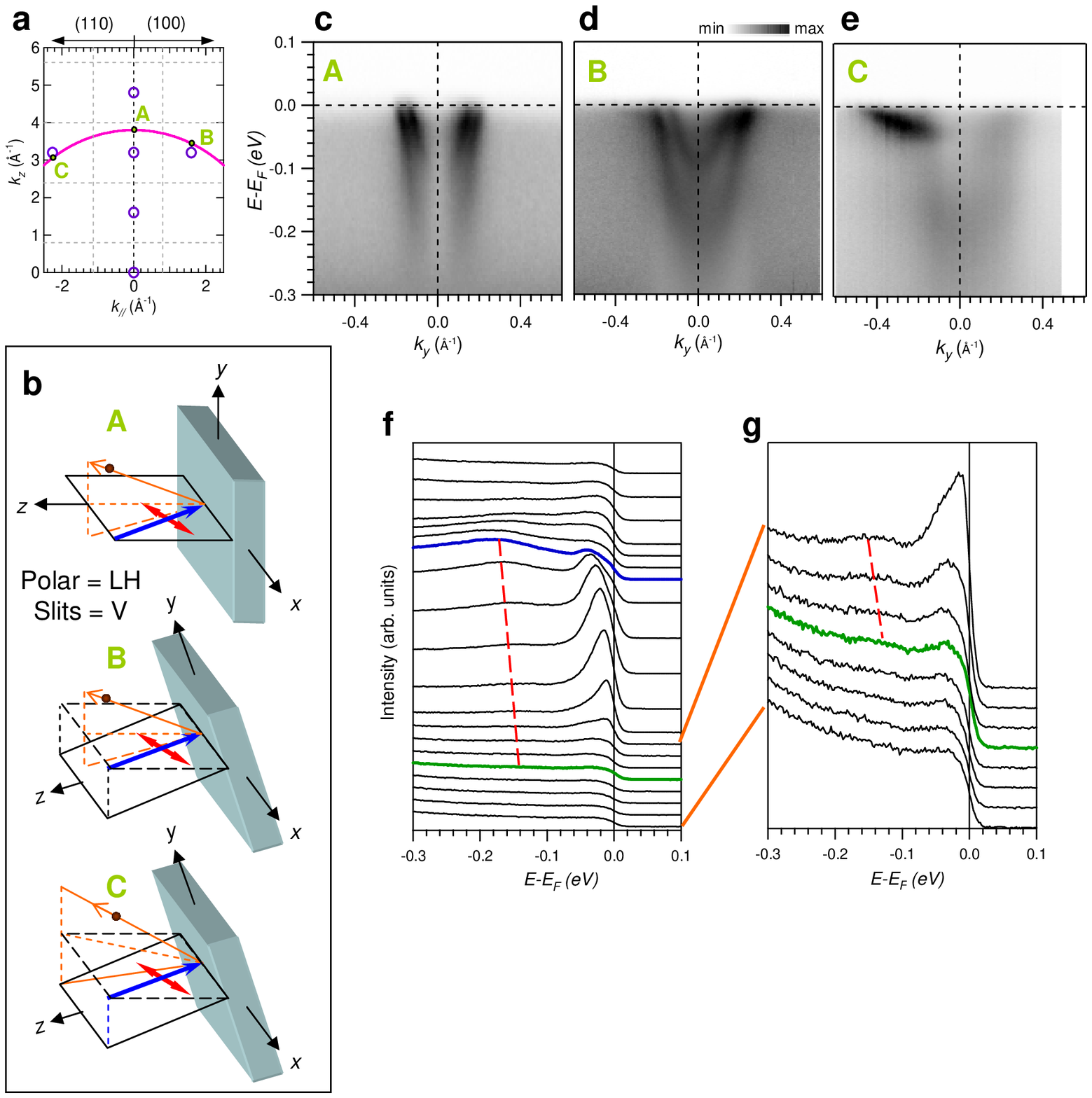}
    \end{center}
    \caption{\label{ANGLE-dependence}
            \footnotesize{
    		{\bf Intensity modulations of the electronic subbands as a function of emission angle.}
    		{\bf a},~Schematic representation of the reciprocal space of SrTiO$_3$ 
    		over several Brillouin zones in the $(100)-(001)$ and $(110)-(001)$ planes. 
    		The continuous line is the spherical shell on which the measurements at $h\nu = 47$~eV are performed.
    		{\bf b},~Schematics of the experimental geometries for the data taken at points
    		A, B and C in this figure.
            {\bf c-e},~Energy-momentum intensity maps near $E_F$ 
            at points A, B and C represented in panel~{\bf a}. 
		    {\bf f-g},~EDCs at point C integrated over intervals of $0.1$~\AA$^{-1}$. The blue EDC is located
		    at $\Gamma$, and the green EDC is at the zone edge
            }
            }
\end{figure}

Fig.~SF~\ref{ANGLE-dependence} shows the energy-momentum intensity maps 
for cuts at three points in reciprocal space
labeled as A, B and C. The spectra at points B and C were already discussed in the main text, 
and are again shown here for clarity and completeness in the discussion.

The three points A, B and C correspond to different positions along the $(001)$ (or $k_z$) axis and different emission angles, 
as represented in Fig.~SF~\ref{ANGLE-dependence}(a)~\cite{Aiura-2002, Takizawa-O2pBands-2009},
such that $\Gamma_{002}\text{A} = 0.7 (\frac{\pi}{a}) = 0.56$~\AA$^{-1}$, 
$\Gamma_{012}\text{B} = 0.25 (\frac{\pi}{a}) = 0.2$~\AA$^{-1}$ and $\Gamma_{112}\text{C} \approx -\Gamma_{012}\text{B}$
($a$ is the cubic SrTiO$_3$ lattice parameter). 
All the spectra in this figure were measured at $h\nu = 47$~eV with LH light polarization 
and detection along $y$ [vertical (V) slits], as schematically shown in Fig.~SF~\ref{ANGLE-dependence}(b). 
At normal emission (point A), the detection plane coincides with the sample's $yz$-plane.
Exactly at $\Gamma$, as explained in the previous section, the only symmetry-allowed states are $d_{xz}$-like,
which have however a very weak intensity at this photon energy.  
Note that in this case the polarization vector has components both along the $x$ and $z$ directions,
the first being odd with respect to the detection plane, and the second even with respect the same plane.
Thus, away from $\Gamma$, at emission along the $yz$-plane, both $d_{xy}$-like bands (odd with respect the $yz$-plane) 
and $d_{yz}$-like bands (even with respect the same plane) can be observed.

On the other hand, for points B-like and C-like, 
the photon polarization is not parallel to any of the sample's symmetry directions or planes,
and the spectra have different symmetry mixtures depending on the measurement point in $k$-space (see also further). 

From figure~SF~\ref{ANGLE-dependence}(c), we see then that at normal emission (point A) 
the two strongly dispersing parabolic bands are distinctly observed. These bands remain
unchanged at points B and C [Figs~SF~\ref{ANGLE-dependence}(d,e)], 
though their spectral intensity decreases as the emission angle increases.
Note, in particular, that although points B and C are equidistant from $\Gamma$, the light parabolic bands
are more intense at point B, where the emission angle is lower. 
Conversely, the two weakly dispersing bands (the `upper and lower shallow bands') 
are best seen at point C, where the emission angle is largest.
Figures~SF~\ref{ANGLE-dependence}(f, g) show the energy distribution curves (EDCs) 
at point C integrated over intervals of $0.1$~\AA$^{-1}$.
The lower shallow band is clearly seen in the raw data dispersing from $\Gamma$ (blue EDC) to the zone edge (green EDC).
Additionally, as noted in the main paper, the same asymmetric enhancement of the negative-$k$ intensities 
for the two shallow bands is observed in point C. This indicates that the two shallow bands have the same symmetry. 
These shallow bands rapidly loose intensity as the emission angle decreases, 
to completely disappear in the spectra at normal emission, further indicating that their symmetry
(or more generally, their spectral function) is not the same of either of the parabolic bands.

From the data of Fig.~SF~\ref{ANGLE-dependence}, it is then apparent that there is a unique underlying band structure at all 
three points A, B and C, further proving the 2D-like character of all the observed bands.
%
%

\begin{figure}
    \begin{center}
    	\includegraphics[width=12cm]{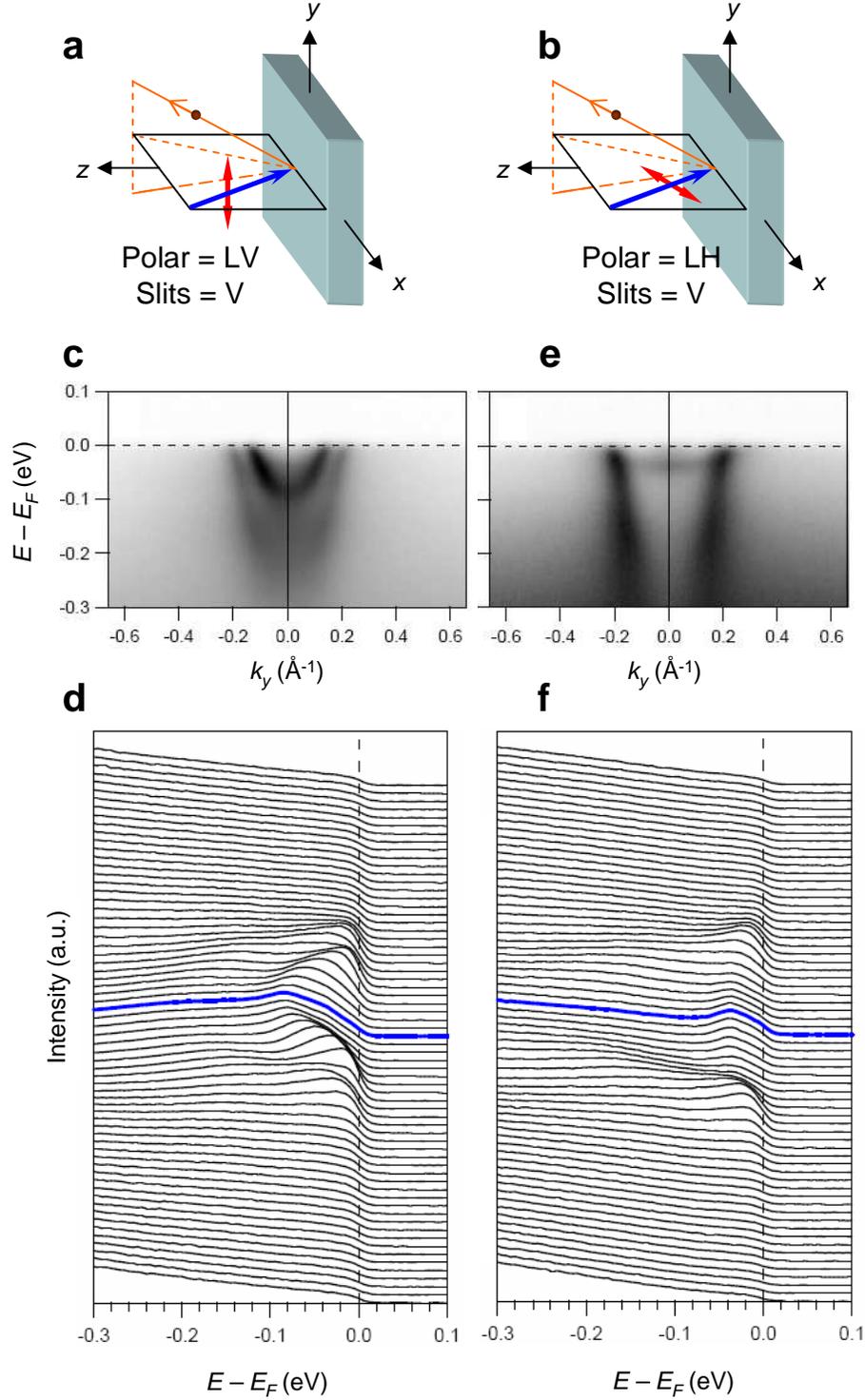}
    \end{center}
    \caption{\label{POLAR-dependence}
    		\footnotesize{
            {\bf Intensity modulations of the electronic subbands as a function of photon polarization.}
    		{\bf a, b},~Schematics of the experimental geometries for the data taken at point B 
    		[Fig.~SF~\ref{ANGLE-dependence}(a)] in this figure.
    		{\bf c, d},~Energy-momentum intensity map and EDCs for LV polarized light.
    		{\bf e, f},~Same measurements for LH polarized light.
    		Blue EDCs are located at $\Gamma_{102}$.
    		Both sets of spectra were obtained in the same spatial spot of the sample, at $T=10$~K.
    		}
    		}
\end{figure}

As described in the main text, the photoemission intensity from the different subbands 
is also strongly modulated by the photon polarization. For the sake of completeness in the discussion,
we show again in Fig.~SF~\ref{POLAR-dependence} the energy-momentum intensity maps and EDCs taken
at $\Gamma_{102}$ (point B) for the non-doped sample.
These data were collected at $h\nu = 45$~eV with V-slits (detection along $y$) using
both LV and LH polarized photons, as shown in Figs.~SF~\ref{POLAR-dependence}(a, b). 
In this geometry, where electrons are detected off-normal emission, 
only the $xz$-plane (coinciding with the light-incidence plane), where $\Gamma_{102}$ lies, is a high-symmetry plane.
For LV polarization, the light electric field is odd with respect to this mirror plane.
Thus, only the $d_{xy}$- and $d_{yz}$-like bands, which are both odd with respect to the $xz$-plane,
can be detected, as indeed observed [Figs.~SF~\ref{POLAR-dependence}(c, d)]. 
On the other hand, for LH polarized light, the electric field
is even with respect to the mirror plane, and only $d_{xz}$-like bands can be detected at $\Gamma_{102}$,
also in agreement with the observation of the upper shallow band [Figs.~SF~\ref{POLAR-dependence}(e, f)]. 
Away from $\Gamma_{102}$ along the detection direction
there is symmetry mixing. Experimentally, the lower parabolic band, but not the upper parabolic band, is again observed.

\subsection{Band bending of the O-$2p$ valence band}

\begin{figure}
    \begin{center}
    	\includegraphics[width=12cm]{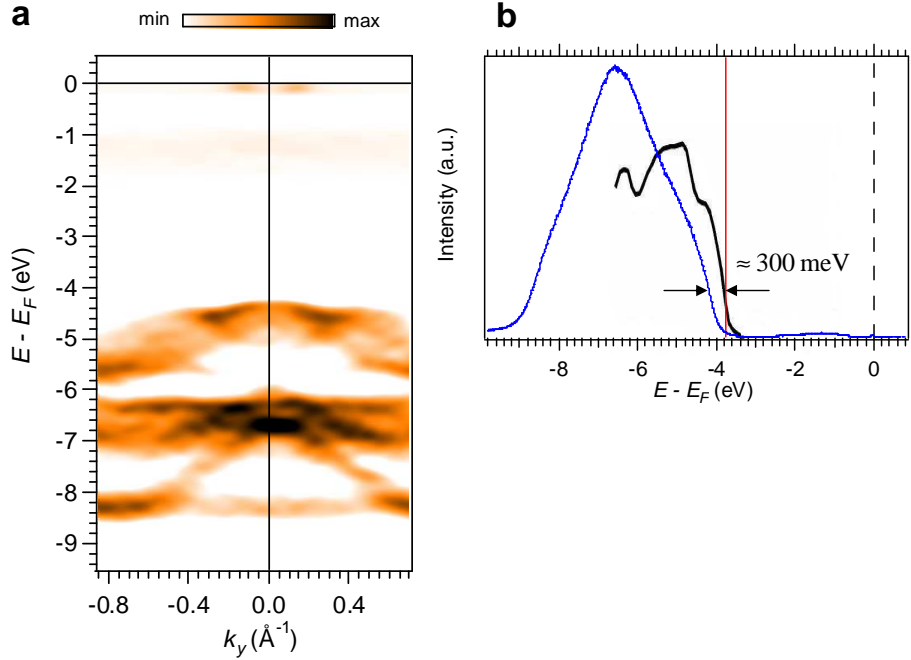}
    \end{center}
    \caption{\label{ValenceBand-O2p}
    		\footnotesize{
            {\bf ARPES spectra and band-bending of the O-$2p$ valence band in SrTiO$_3$.}
    		{\bf a},~Second derivative of the energy-momentum intensity map over a wide energy range,
    		showing the dispersion of the O-$2p$ valence band at energies between $4 - 9$~eV. 
    		The non dispersive oxygen-vacancy-induced intra-gap states at $-1.3$~eV, 
    		and the QP band ar $E_F$, are also observed.
    		{\bf b},~Raw EDC at normal emission ($k_y = 0$) from the data in {\bf a} (blue curve).
    		The energy-dependent optical extinction coefficient measured in Ref.~\cite{Ellipsometry-STO-2001}
    		is shown for comparison (black curve). The leading edge of the photoemission spectrum is shifted down by about
    		300~meV with respect to the optical absorption leading edge.
    		}
    		}
\end{figure}

Figure~SF~\ref{ValenceBand-O2p}(a) shows the angle-resolved data over a wide energy range at normal emission,
LH polarization, V-slits and $h\nu= 47$~eV [point A in figure~SF~\ref{POLAR-dependence}(a)]
for the sample that has bulk doping $n_{3D} = 10^{18}$~cm$^{-3}$.
The dispersion of the oxygen valence band, roughly located in between -9~eV and -4~eV, is clearly observed, 
attesting the high-quality of the surfaces obtained after cleaving. 
As seen from figures~SF~\ref{ValenceBand-O2p}(a, b), the peak position of the valence-band maximum (VBM)
is located at $E \approx -4.35$~eV, and the leading-edge of the valence band is located at about $-3.9$~eV. 
These values are systematically lower by 200-500~meV than the peak position of the bulk VBM,
predicted to be about $-3.9$~eV by tight-binding calculations~\cite{Takizawa-O2pBands-2009},
and the bulk optical band gap of 3.75~eV~\cite{Ellipsometry-STO-2001}, shown for comparison.
These results are consistent with the dispersions and shifts 
reported in previous works~\cite{Aiura-2002, Takizawa-O2pBands-2009}.
More important, the observed energy down-shift of the oxygen valence band is in quantitative
agreement with the confining potential at the surface, estimated to be $V_0 \approx -300$~meV (see below and main paper),
and responsible for the formation of the two-dimensional electron gas and the subbands beneath the surface
of cleaved SrTiO$_3$. This indicates that such a band shift of the O-$2p$ valence band is due to band-bending 
by the above confining potential.
Furthermore, as noted in Ref.~\cite{Aiura-2002}, when the cleaved surface is exposed to oxygen, 
the O-$2p$ valence band shifts back by about 200~meV towards $E_F$, lending support
to the hypothesis that the confining potential, in the case of cleaved SrTiO$_3$, is due to
surface oxygen vacancies. 

\subsection{Wedge model}

We approximate the effect of the confining potential acting in the $z$-direction
by a potential wedge $V(z) = V_0 +  e F z$, where $e$ is the charge of
the electron and $F$ is the strength of the electric field along $z$. 
The quantized eigenenergies $E_n$ are given in very good approximation by~\cite{Ueno-Electric-Field-Induced-SC-Plastic-STO-2008, 
Semicond-2DEGs-WedgeWell-1982}: 
\begin{equation}
	E_n = V_0 + \left( \frac{\hbar^2}{2m^{\star}_{z}} \right)^{1/3} 
				\left[ \left( \frac{3\pi}{2} \right) \left( n - \frac{1}{4} \right) e F \right]^{2/3},  
\end{equation}
where $n = 1 {\text , } 2 {\text , } 3 {\text , } {\dots}$, and $m^{\star}_{z}$ is the effective mass in the $z$-direction.
The exact eigenenergies have $n - \frac{1}{4}$ in this equation replaced by 0.7587, 1.7540, and 2.7575,
respectively, for the three lowest solutions~\cite{Semicond-2DEGs-WedgeWell-1982}. 
The average value of $z$ for the $n$-th subband is 
${\rm e} F \left< z_n \right> = \frac{2}{3} |E_n - V_0|$~\cite{Semicond-2DEGs-WedgeWell-1982}.
  
Replacing with the appropriate physical constants, one gets
\begin{equation}
	E_n = V_0 + 9 \times 10^{-7} \left( \frac{m_e}{m^{\star}_{z}} \right)^{1/3} 
				\left( n - \frac{1}{4} \right)^{2/3} F^{2/3}, 
\end{equation}
where $m_e$ is the free electron mass, $E_n$ and $V_0$ are expressed in eV, and $F$ is expressed in V/m.
This expression allows to determine the field strength $F$ from
the experimental subband splitting $\Delta E = E_2 - E_1$ 
between the first to subbands of $d_{xz/yz}$ character. 
Notice that, in cubic symmetry, the $d_{xz}$- and $d_{yz}$-like bands have the 
same small effective mass in the $z$-direction, with $m^{\star}_{z} \approx 0.7 m_e$ from our experiments.

From the experimental value $\Delta E \approx 120$~meV, we get
$F \approx 83$~MV/m.

Given the wedge-like form of the potential, the states with higher energy 
will have the largest extension along $z$. 
In our case, the observed highest occupied state is the upper shallow subband,
which would correspond to the $n=2$ state of the $d_{xz}$-like band.
Thus, with the above field strength, we can estimate the depth ($L$) of the confining
potential from the approximation $e F L \approx  [E_2 (d_{xz}) - E_1 (d_{xz}]$ 
(note that $V_0$ is not needed to compute this). This gives $L \approx 14.5$~\AA,
or about 3.7 unit cells.

The potential at the bottom of the wedge well ($V_0$) can, in turn, be estimated from the field strength and 
the bottom of the $d_{xy}$ parabolic band [$E_1 (d_{xy}) \approx -210$~meV], which has a heavy mass along $z$ of
$m^{\star}_{z} \approx 20 m_e$. This yields $V_0 \approx -260$~meV.  This value allows an independent
estimate of the width of the well from the average value of $z$ for the $n=2$ subband, namely 
${\rm e} F \left< z_2 \right> = \frac{2}{3} |E_2 - V_0|$.
Using $E_2(d_{xz}) = -40$~meV, we obtain $\left< z_2 \right> = 17.7$~\AA, or about $4.5$~unit cells.
Additionally, as the potential $V(z)$ remains negative ({\it i.e.}, confining) 
while $e F z \lesssim V_0$, this yields $L_{max} \approx 31$~\AA~$\approx 8$~unit cells 
as a maximum bound for the width of the 2DEG. 
These independent estimates provide important crosschecks for the internal consistency of our analysis.

The energy of the bottom of the different $d_{yz/xz}$ subbands is then readily calculated
from their light effective masses along $z$ and the values of $V_0$ and $F$.
One obtains: $E_1 (d_{yz/xz}) = -100$~meV, and $E_2 (d_{yz/xz}) = +20$~meV.
This agrees with the observations: only the first $d_{yz}$ subband (the small light parabola in the $xy$-plane)
lies below $E_F$. The experimentally observed degeneracy lift between the $d_{xz}$ and $d_{yz}$ bands
is not taken into account by this simple model.  From the data, we know that the $E_1(d_{xz})$ subband
(the lower shallow subband) is shifted by about 60~meV below its $d_{yz}$ partner. One then expects
that the upper shallow subband lies at an energy
$E_2(d_{xz}) = E_2 (d_{xz/yz}) - 60{\rm meV} = -40$~meV, again in excellent agreement with the observations.

Finally, we can also use our estimate of $F$ to compute the superficial charge density $\sigma_{2D}$
induced by the confining potential.
We follow the same type of calculation used by Ueno and coworkers 
(Ref.~\cite{Ueno-Electric-Field-Induced-SC-Plastic-STO-2008}, Supplementary Material), namely:
\begin{equation}
	\frac{e}{2} \sigma_{2D} = \int_{0}^{F}\epsilon_0 \epsilon(F^{\prime}) dF^{\prime},
\end{equation}
where $\epsilon_0$ is the vacuum dielectric constant and $\epsilon(F)$ is the material's dielectric permittivity.
Empirically, it is well established that the dielectric permittivity of SrTiO$_3$ depends on the
electric field strength $F$ according to the relation~\cite{Neville-Epsilon-vs-F-STO}: 
$\epsilon(F) \approx 1/(4 \times 10^{-5} + 5 \times 10^{-10} F)$, where $F$ is in V/m.
We then readily obtain $\sigma \approx 0.25 {\rm ~} e/a^2$,
where $a = 3.905$~\AA~is the cubic lattice parameter.
This value is consistent with the experimental electron count $\sim 0.31 - 0.36{\rm ~} e/a^2$,
obtained from the area of the Fermi surfaces (see main text).

\subsection{LDA slab calculations of the electronic structure in the presence of surface oxygen vacancies}

\begin{figure}
    \begin{center}
    	\includegraphics[width=12cm]{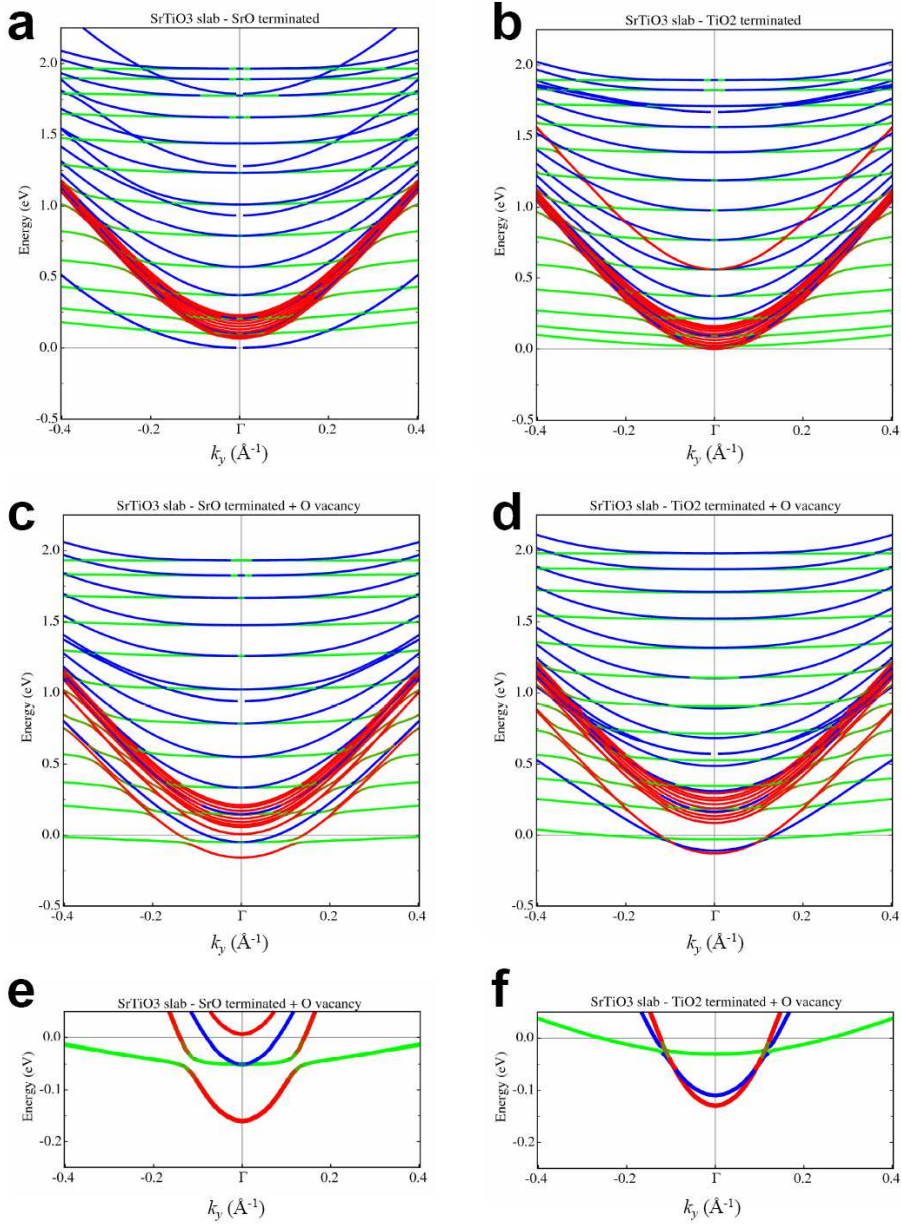}
    \end{center}
    \caption{\label{LDA-slabs-OVacs}
    		\footnotesize{
            {\bf Band structure calculation of $1 \times 1 \times 12$ slabs of cubic SrTiO$_3$ 
             with one oxygen vacancy on the surface.}
             {\bf a, b} Calculations for SrO and TiO$_2$ terminated surfaces, respectively. 
    		 A subband structure of levels is created due to the $z$-direction confinement, 
    		 with a large level spacing for the $d_{xz/yz}$ doublets, which share the same light mass $m^{\star}_{z}$. 
    		 In contrast the $d_{xy}$ levels are hardly split due to their large $m^{\star}_{z}$. 
    		 Panels {\bf c} and {\bf d} include one oxygen vacancy defect at the SrO and TiO$_2$ surfaces. 
    		 One clearly observes the double effect of doping and confinement. 
    		 The bottom of the levels is pulled beneath the Fermi energy. Panels {\bf e} and {\bf f}
    		 show an enlarged detail of the energy region near $E_F$. 
    		 Colors indicate the character of each band along $k_y$, following the same convention
    		 as in the main paper: $d_{xy}$-like bands are red, $d_{yz}$-like bands are blue, 
    		 and $d_{xz}$-like bands are green. 
    		}
    		}
\end{figure}

To illustrate the effects on the band structure of SrTiO$_3$ due to the presence of oxygen vacancies  near the surface,
we performed {\it ab initio} density functional calculations within the local density approximation. 
We adopted the Wien2k code and modelled the surface by considering slabs of $1 \times 1 \times 12$ cells. 
We considered both SrO and TiO$_2$ terminations, and put one oxygen vacancy defect at the surface. 
Our goal here is to understand, through a realistic yet computationally simple calculation, 
whether the presence of such vacancies can yield a 2DEG at the surface 
with the concomitant energy lowering for the slab states. 
Thus, in this simple initial study, we considered the undistorted cubic cell and
did not relax the lattice. Our results are presented in Fig.~SF~\ref{LDA-slabs-OVacs}. 
They show that the main features of the 2DEG
at the surface of vacuum-cleaved SrTiO$_3$, discussed in the main text, are borne out by this calculation. 
Namely, that the bottom of the $t_{2g}$ manifold around the $\Gamma$-point is lowered in energy 
by the effective attractive potential by as much as 200 meV. 
Moreover, the results also correctly capture the fact that the $d_{xy}$ parabolic band, 
due to its large effective mass along the $z$-direction, is pulled down more that the $d_{xz/yz}$ doublet. 

While it is appealing that these generic features are well captured by the LDA results, one should bear in mind
that due to the technical limitations in the cell size that can be considered, our calculations remain on the qualitative 
level.
A larger computational effort, including clusters of surface vacancies, is in progress.

